\documentclass[twocolumn,showpacs,floats,floatfix,superscriptaddress,aps,pra]{revtex4-1}
\usepackage{amsfonts}
\usepackage{amssymb}
\usepackage{amsmath}
\usepackage{calc}
\usepackage{graphicx}
\usepackage{bm}

\usepackage[normalem]{ulem}
\usepackage{amsmath,amssymb}
\usepackage{multirow}
\usepackage{xcolor,soul}
\usepackage{srcltx}
\usepackage{hyperref,graphicx}

\begin{document}

\author{S. Longhi} 
\email{stefano.longhi@polimi.it}
\affiliation{Dipartimento di Fisica, Politecnico di Milano and Istituto di Fotonica e Nanotecnologie del Consiglio Nazionale delle Ricerche, Piazza L. da Vinci 32, I-20133 Milano, Italy}
\affiliation{IFISC (UIB-CSIC), Instituto de Fisica Interdisciplinar y Sistemas Complejos - Palma de Mallorca, Spain}

\title{Non-Bloch band collapse and chiral Zener tunneling}
  \normalsize


%
\bigskip
\begin{abstract}
\noindent  
Non-Bloch band theory describes bulk energy spectra and topological invariants in non-Hermitian crystals with open boundaries, where the bulk eigenstates are squeezed toward the edges (skin effect). However, the interplay of non-Bloch band theory, skin effect and coherent Bloch dynamics is so far unexplored. In two-band non-Hermitian lattices, it is shown here that collapse of non-Bloch bands and skin modes deeply changes the Bloch dynamics under an external force. In particular, for resonance forcing non-Bloch band collapse results in Wannier-Stark ladder coalescence and chiral Zener tunneling between the two dispersive Bloch bands.
\end{abstract}

\maketitle

{ \it Introduction.}
Bloch band theory is the fundamental tool to describe electronic states in crystals \cite{r1}. Under the action of a weak dc electric field, Bloch theory predicts that electrons undergo an oscillatory motion,  the famous Bloch oscillations (BOs) \cite{r1}, which can be explained from the formation of a Wannier-Stark (WS) ladder energy spectrum \cite{Wannier}. Transitions among different bands occur for stronger fields because of Zener tunneling (ZT) \cite{r2}. BOs and ZT are ubiquitous phenomena of coherent wave transport in periodic media and have been observed in a wide variety of physical systems  \cite{BO1,BO2,BO3,BO4,BO5,BO6,BO7,BO8,BO9,BO10,BO11}. 
It is remarkable that, after one century from the seminal paper by Felix Bloch \cite{r1}, there are still treasures to be uncovered within Bloch band theory. For example, Bloch band theory is central in the understanding  of topological insulators \cite{r3,r4,r5} and in the description of flat band systems showing unconventional localization, anomalous phases, and strongly correlated
states of matter  \cite{r6,r7,r8,r9,r10,r11,r11b,r12,r13,r14,r15}. Recently, a great interest is devoted to extend Bloch band theory and topological order to non-Hermitian lattices \cite{r17,r18,r19,r20,r21,r22,r23,r24,r25,r26,r27,r28,r29,r30,r31,r32,r33,r34,r35,r36,r37,r38,r39,r40,r41,r42,r43,r44,r44b,r45,r46,r47,r48,r49,r50,r51,r51b,r51c,r51d}.
 In Bloch band theory,  an electronic state is  defined by the quasi-momentum $\mathbf{k}$, which spans the first Brillouin zone, and is delocalized all along the crystal, regardless periodic boundary conditions (PBC) or open boundary conditions (OBC) are assumed. However, in non-Hermitian crystals something strange happens: energy bands in crystals with OBC are described by non-Bloch bands that deviate from ordinary Bloch bands; bulk eigenstates cease to be delocalized and get squeezed toward the edges (non-Hermitian skin effect); and the bulk-boundary correspondence based on Bloch topological invariants generally fails to correctly predict the existence of topological zero-energy modes \cite{r16,r20,r21,r22,r23,r24,r25,r26,r32,r43,r45}. Recent seminal works \cite{r22,r24,r32} showed that to correctly describe energy spectra and topological invariants in crystals with OBC one needs to extend Bloch band theory so as the quasi-momentum $\mathbf{k}$ becomes complex and varies on a generalized Brillouin zone. Bloch and non-Bloch bands can  show different symmetry breaking phase transitions and, interestingly, band flattening near an exceptional point (EP) can be observed for non-Bloch bands, while ordinary Bloch bands remain dispersive \cite{r25,r43,r51b}. As  BOs and ZT in non-Hermitian lattices have been investigated in some recent works \cite{r52,r53,r54,r55,r56,r57}, the implications of non-Bloch band theory to particle Bloch dynamics remain obscure.\\ 
 In this Letter we consider Bloch dynamics in a two-band non-Hermitian crystal and unveil  a dynamical behavior unique to non-Hermitian crystals: ZT between dispersive Bloch bands, induced by resonance forcing, becomes chiral at the non-Bloch band collapse point, i.e. electrons irreversibly tunnel from one dispersive Bloch band to the other one, contrary to Hermitian systems where ZT is reversible (oscillatory). Chirality of ZT can be regarded as the signature of non-Bloch band collapse.  \par
{\it Non-Bloch band collapse.}
We consider a one-dimensional tight-binding lattice with chiral symmetry and with two sites per unit cell, driven by an external dc force $F$. In the Wannier basis representation, the Schr\"odinger equation describing the evolution of the occupation amplitudes $a_n$ and $b_n$ in the two sublattices A and B at the $n$-th unit cell reads
\begin{eqnarray}
i \frac{da_n}{dt} & = & \sum_{l=-q}^{q} \rho_{l} a_{n-l}+ \sum_{l=-q}^{q}  \theta_{l} b_{n-l} -Fn a_n  \\
i \frac{db_n}{dt} & = & \sum_{l=-q}^{q}  \varphi_{l} a_{n-l}- \sum_{l=-q}^{q} \rho_{l} b_{n-l}-F n b_n.
\end{eqnarray}
where $\rho_n$, $\theta_n$, and $\varphi_n$ describe the hopping amplitudes and where we assume that electrons can hop up to the $q$-th nearest unit cells. A Hermitian crystal corresponds to $\rho_{-n}=\rho_n^*$ and $\theta_{-n}=\varphi_{n}^*$.\\ 
For an undriven lattice ($F=0$), the Hamiltonian in Bloch basis representation reads 
\begin{equation}
H(k)=\sigma_x d_x(k)+\sigma_y d_y(k)+ \sigma_z d_z(k), 
\end{equation}
 where $\sigma_{x,y,z}$ are the Pauli matrices and $d_x(k)  \equiv (1/2) \sum_n (\theta_n+\varphi_n) \exp(-ikn)$, 
$d_y(k) \equiv (1/2i) \sum_n (\varphi_n-\theta_n) \exp(-ikn)$,
$d_z(k)  \equiv \sum_n \rho_n \exp(-ikn)$.
The energy spectrum shows chiral symmetry $E \leftrightarrow -E$ with the dispersion curves of the two bands given by 
\begin{equation}
E_{\pm}= \pm \sqrt{Q(\beta)}, 
\end{equation}
where $\beta \equiv \exp(-ik)$ and $Q(\beta) \equiv d_x^2+d_y^2+d_z^2$. Note that $Q(\beta)$ is given by a sum of powers of $\beta$, namely
\begin{equation}
Q(\beta)= \sum_{l,n=-q}^{q} (\rho_{l} \rho_{n}+\theta_l \varphi_n) \beta^{n+l}=P_{2q}(\beta)+\frac{R_{2q-1}(\beta)}{\beta^{2q}} 
\end{equation}
where $P_{2q}$ and $R_{2q-1}$ are two polynomials of $\beta$ of order $2q$ and $2q-1$, respectively.
Owing to the non-Hermitian skin effect, the energy bands and corresponding eigenfunctions for a long crystal, comprising $N$ unit cells, are different for PBC and OBC \cite{r22,r32}. For PBC, $k$ spans the first Brillouin zone $-\pi \leq k < \pi$, $\beta= \exp(-ik)$ varies on the unit circle $C_{\beta}$ in complex plane, i.e. $|\beta|=1$, and the energy curves (4) describe the ordinary lattice Bloch bands.  We assume that the two Bloch bands are separable and $Q(\beta) \neq 0$ in the interior of $C_{\beta}$, so that $E_{\pm}(k+ 2\pi)=E_{\pm}(k)$. 
For a lattice with OBC, the energy spectrum is obtained again from Eqs.(4) and (5) but with $\beta$ varying on a generalized Brillouin zone $\tilde{C}_{\beta}$ in complex plane \cite{r22,r32}. The corresponding curves $E_{\pm}(\beta)$ describe so-called non-Bloch bands, and the bulk eigenstates are squeezed toward the left or right lattice edges depending on whether $| \beta|>1$ or $|\beta|<1$. The generalized Brillouin zone $\tilde{C}_{\beta}$ and corresponding non-Bloch energy bands are obtained from the implicit relation \cite{r32}
\begin{equation}
|\beta_{2q}(E)|=|\beta_{2q+1}(E)|
\end{equation}
where, for a given complex energy $E$, $\beta_l(E)$ are the roots of the algebraic equation 
\begin{equation}
\beta^{2q} \left\{ P_{2q}(\beta)-E^2 \right\}+R_{2q-1}(\beta)=0
\end{equation}
ordered such that $|\beta_1(E)| \leq |\beta_2(E)| \leq ... \leq |\beta_{4q}(E)|$ \cite{r32}. A remarkable property of non-Hermitian lattices is that one can observe non-Bloch flat bands, i.e. $Q(\beta)$ independent of $\beta$ as $\beta$ varies on the generalized Brillouin zone $\tilde{C}_{\beta}$, while the ordinary Bloch bands remain dispersive, i.e. $Q(\beta)$ is not constant when $\beta$ varies on the unit circle $C_{\beta}$.  In particular, it can be shown \cite{r58} that, whenever the hopping terms $\rho_n$, $\theta_n$ and $\varphi_n$ become small and vanish for any $n<0$ (or likewise for any $n>0$), the two non-Bloch bands become flat around the two values $E_{\pm}= \pm E_0$, where we have set
\begin{equation}
E_0 \equiv \sqrt{\rho_0^2+\varphi_0 \theta_0}.
\end{equation}
\begin{figure}[htbp]
  \includegraphics[width=87mm]{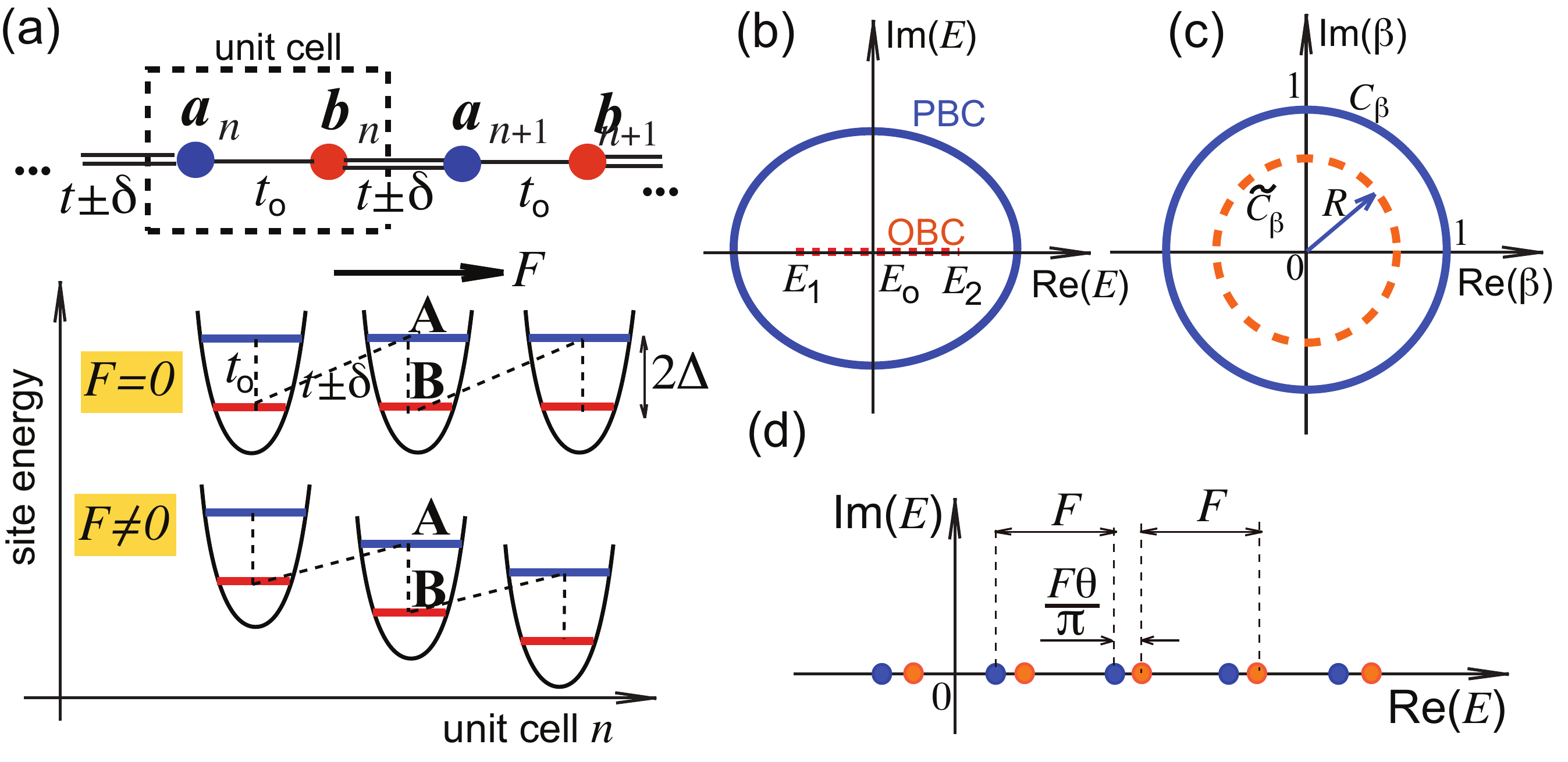}\\
   \caption{(color online) (a) Example of non-Hermitian lattice showing skin effect. The lattice is Hermitian for $\delta=0$. As $\delta$ is increased, non-Bloch bands collapse at $\delta=t$. An external force $F$  introduces a gradient in the site energies (lower panel). (b) Bloch and non-Bloch lattice bands in complex energy plane for $F=0$. The PBC energy spectrum (solid curve) is a closed loop around $E_0=\sqrt{\Delta^2+t_0^2}$, whereas the OBC energy spectrum (dashed curve) is the segment $(E_1,E_2)$ on the real energy axis, with $E_{1,2}= \sqrt{\Delta^2+ (t_0 \mp \sqrt{t^2-\delta^2})^2}$. The OBC band collapses to $E_0$ at $\delta=\pm t$. Only the positive energy branch $E_+(k)$ is shown. (c) Diagram of the Brillouin zone $C_{\beta}$ (unit circle) and generalized Brillouin zone $\tilde{C}_{\beta}$. $\tilde{C}_{\beta}$ is the circle of radius $R$ which shrinks to $\beta=0$ (skin mode collapse) as $\delta\rightarrow t$. (d) Energy spectrum for a non vanishing force. The spectrum is entirely discrete (point spectrum) and formed by two interleaved WS ladders.} 
\end{figure}
Correspondingly, the generalized Brillouin zone $\tilde{C}_{\beta}$ shrinks toward the point $\beta=0$ (or $1 / \beta=0$), i.e. the collapse of the non-Bloch bands is associated with the coalescence of the skin modes, indicating that at the band collapse the OBC Hamiltonian has two high-order EPs \cite{r58}.  An example of non-Bloch band collapse is shown in Fig.1 for a lattice model which is a variant of the Hamiltonian earlier introduced in Ref.\cite{r22} (see also \cite{r16,r25,r43}). The model corresponds to the following non-vanishing values of couplings [Fig.1(a)]: $\rho_0=\Delta$, $\theta_0=\varphi_0=t_0$, $\theta_1=t+\delta$, and $\varphi_{-1}=t-\delta$, with $\Delta$, $t_0$, $t$ and $\delta$ real positive numbers. The lattice is Hermitian for $\delta=0$. As $ \delta $ is increased, a non-Bloch band collapse at the energies $\pm E_0= \pm \sqrt{\Delta^2+t_0^2}$ occurs at $\delta=  t$. The generalized Brillouin zone $\tilde{C}_{\beta}$ for this model is a circle of radius $R=\sqrt{(t-\delta)/(t+\delta)}$, which shrinks as $\delta \rightarrow t$ [Fig.1(c)]. 
Note that, at the non-Bloch bands collapse point, the ordinary Bloch bands remain dispersive, i.e. band flattening occurs for non-Bloch bands solely, which is a distinctive feature than Bloch band flatting in non-Hermitian systems studied in some recent works \cite{r59,r60,r61,r62}. \par
{\it Chiral Zener Tunneling between dispersive Bloch bands.}  Let us now consider the case of a driven lattice, i.e. $F \neq 0$. 
Like for an Hermitian lattice \cite{r62b,r62c}, the external force changes the energy spectrum from an absolutely continuous spectrum (the two energy bands) into a point spectrum composed by two interleaved  WS ladders \cite{r58}; see Fig.1(d). Owing to the formation of the WS ladders, the eigenstates become localized in the bulk. Hence in the driven non-Hermitian lattice the distinction between PBC and OBC smears out and one can disregard boundary conditions assuming that excitation in the lattice remains confined far from the edges. To study ZT between the two dispersive Bloch bands, let us consider wave dynamics in the Bloch basis representation. After setting $a_n(t)=\int_{-\pi}^{\pi} dk A(k,t) \exp(ikn)$ and  $b_n(t)=\int_{-\pi}^{\pi} dk B(k,t) \exp(ikn)$, the evolution equation for the spectral amplitudes  $A$ and $B$ reads
\begin{equation}
i \frac{\partial}{\partial t}
\left(
\begin{array}{c}
A \\
B
\end{array}
\right)= \left( H(k)-iF \frac{\partial}{\partial k} \right)
\left(
\begin{array}{c}
A \\
B
\end{array}
\right).
\end{equation}
The energy spectrum $E^{(WS)}$ of the two interleaved WS ladders can be calculated from Eq.(9) by standard methods  and reads  (technical details are given in \cite{r58})
\begin{equation}
E^{(WS)}_{\pm}=l F \pm \frac{F \theta}{2 \pi}
\end{equation}
($l=0, \pm1, \pm2,...$), where $\theta=\theta(F)$ is a complex angle such that $\cos \theta$ is the half trace of the $ 2 \times 2$ ordered exponential matrix 
\begin{equation}
U= \int_{-\pi}^{ \pi} dk \exp \left\{ -i H(k) /F \right\}.
\end{equation}
Here we are mainly interested in the case where the angle $\theta$ is real, so that the energy spectrum $E^{(WS)}_{\pm}$ is real (despite the Hamiltonian is not Hermitian). This case occurs rather generally when $\rho_0$ is a real number  (positive for the sake of definiteness) and the couplings $|\rho_l|$ ($l \neq 0$), $|\varphi_l|$ and $|\theta_l|$  are much smaller than $\rho_0$. In this limiting case the two weakly-dispersive Bloch bands of the lattice describe closed loops around $ \pm E_0 \simeq \pm \rho_0$ and correspond to particle occupation mostly in the sublattice A (for the  the $^{\prime}$higher$^{\prime}$ energy band $E_+$) and in sublattice B (for the other $^{\prime}$lower$^{\prime}$ energy band $E_-$). In the weak forcing limit $F \rightarrow 0$, an approximate expression of $\theta$ can be derived by a standard WKB analysis and reads 
\begin{equation}
\theta \simeq 	\frac{1}{F} \int_{-\pi}^{\pi} dk E_{+}(k)=\frac{2 \pi}{F} 	\sqrt{ \sum_l ( \rho_l \rho_{-l}+\varphi_l \theta_{-l} )}.
\end{equation}
As shown below such a result turns out to be {\em exact} for any strength $F$ of forcing at the non-Bloch band collapse, i.e. when $\rho_l=\theta_l=\varphi_l=0$ for $l<0$ (right unidirectional hopping) or for $l>0$ (left unidirectional hopping), so as $\theta= 2 \pi E_0/F$. Figure 2 shows an example of the numerically-computed behavior of the angle $\theta$ versus $F$ for the lattice model of Fig.1(a) and for increasing values of parameter $\delta$ until non-Bloch band collapse is attained. The dashed curves in the figure correspond to the approximate result obtained from the WKB analysis. Note that at non-Bloch band collapse point the WKB analysis becomes exact [Fig.2(c)]. An interesting case is observed when $\theta=0, \pi$, corresponding to the coalescence of the two WS ladders $E^{(WS)}_+$ and $E^{(WS)}_-$. At the non-Bloch band collapse point such a coalescence is attained for resonance forcing 
\begin{equation}
F= \pm 2E_0/n
\end{equation}
\begin{figure}[htbp]
  \includegraphics[width=84mm]{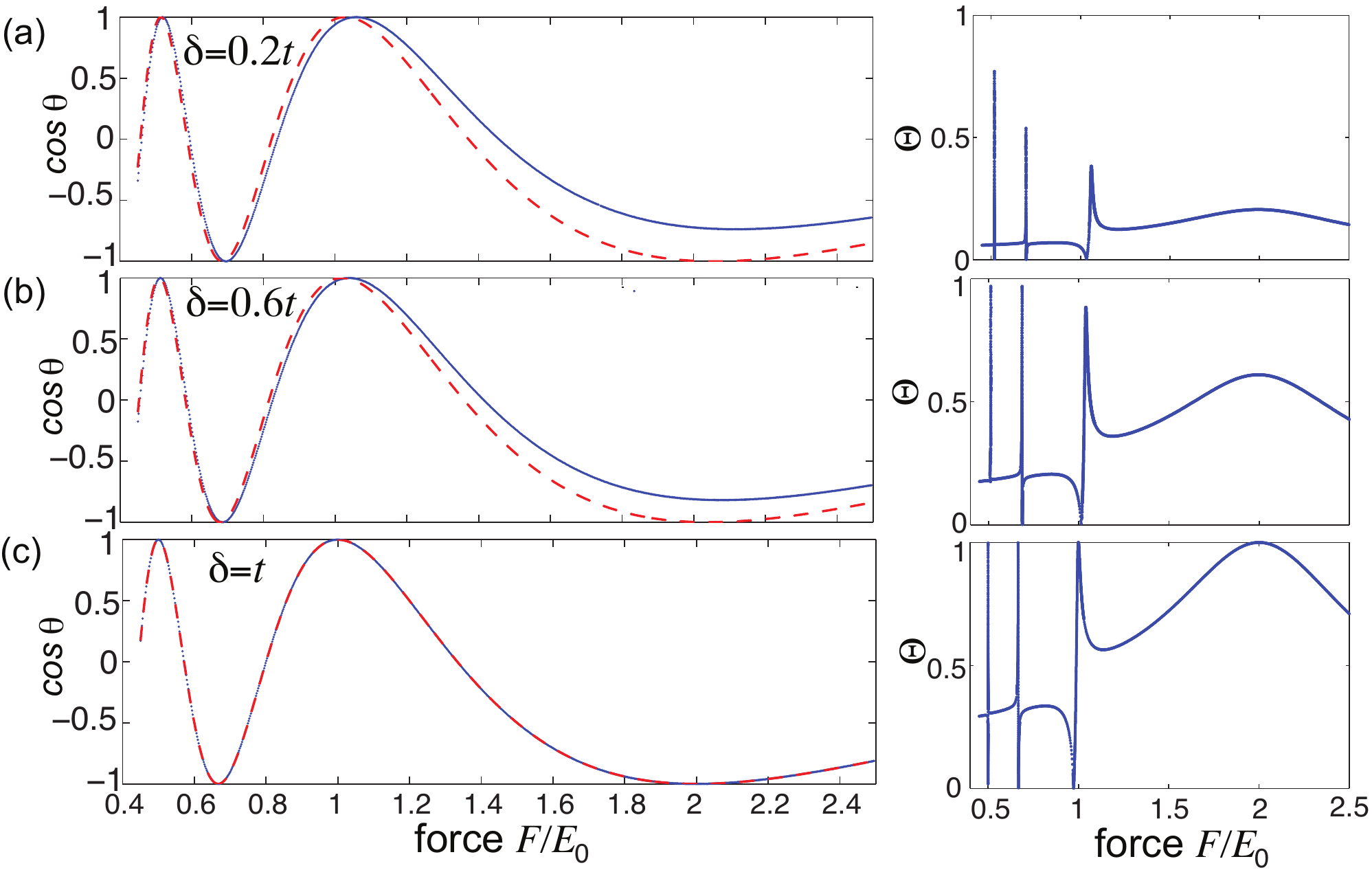}\\
   \caption{(color online) Left panels: Behavior of $\cos \theta$ versus normalized force $F/E_0$ in the driven lattice of Fig.1 for a few increasing values of $\delta /t$: (a) $\delta /t=0.2$, (b) $\delta /  t=0.6$, (c) $\delta /  t=1$ (non-Bloch band collapse). Other parameters are $\Delta /t=2$, $t_0 / t=0.4$. The dashed curves show the behavior of $\cos \theta$ versus $F/E_0$ predicted by the WKB analysis. Right panels: behavior of the scalar product $\Theta$ versus $F/E_0$.}
\end{figure}
($n=1,2,3,...$). While in the Hermitian case such an accidental coalescence yields exact wave packet revivals at time intervals multiplies than $t_B=2\pi /F$ \cite{r62b,r62c}, in the non-Hermitian lattice the coalescence of energy ladders can correspond to the simultaneous coalescence of the localized WS eigenstates, i.e. to a WS EP \cite{r63}. This occurs when the eigenvectors $\mathbf{u}_{1,2}$ of the matrix $U$ [Eq.(11)]  become parallel, i.e. when the scalar product $\Theta=|\langle \mathbf{u}_1| \mathbf{u}_2 \rangle | / \sqrt{\langle \mathbf{u}_1| \mathbf{u}_1 \rangle \langle \mathbf{u}_2| \mathbf{u}_2 \rangle }$ becomes equal to one (right panels in Fig.2).  The main result of this work is that (i) a WS EP is observed when the non-Bloch bands collapse and the external force satisfies the resonance condition (13); (ii) at a WS EP, ZT between the two dispersive Bloch bands becomes chiral, i.e. it ceases to be oscillatory and irreversible tunneling from one dispersive band to the other one is observed. To prove such a statement, let us notice that, according the $^{\prime}$acceleration theorem$^{\prime}$ \cite{r62c,r64,r65}, the external force induces a drift of the Bloch wave number $k$ in time according to $k=k_0+Ft$ ($k_0$ is the particle quasi-momentum at initial time) to cyclically span the Brillouin zone $C_{\beta}$. We then look for an {\em exact} solution to Eq.(9) of the form  $A(k,t)=f_A(t) \delta(k-k_0-Ft)$,  $B(k,t)=f_A(t) \delta(k-k_0-Ft)$, and assume without loss of generality $k_0=0$. The evolution equation for the amplitudes $f_A(t)$ and $f_B(t)$ to find the particle in either sublattices A and B reads
\begin{equation}
i \frac{d}{dt} 
\left(
\begin{array}{c}
f_A \\
f_B
\end{array}
\right)=H(k=Ft) 
\left(
\begin{array}{c}
f_A \\
f_B
\end{array}
\right).
\end{equation}
\begin{figure}[htbp]
  \includegraphics[width=84mm]{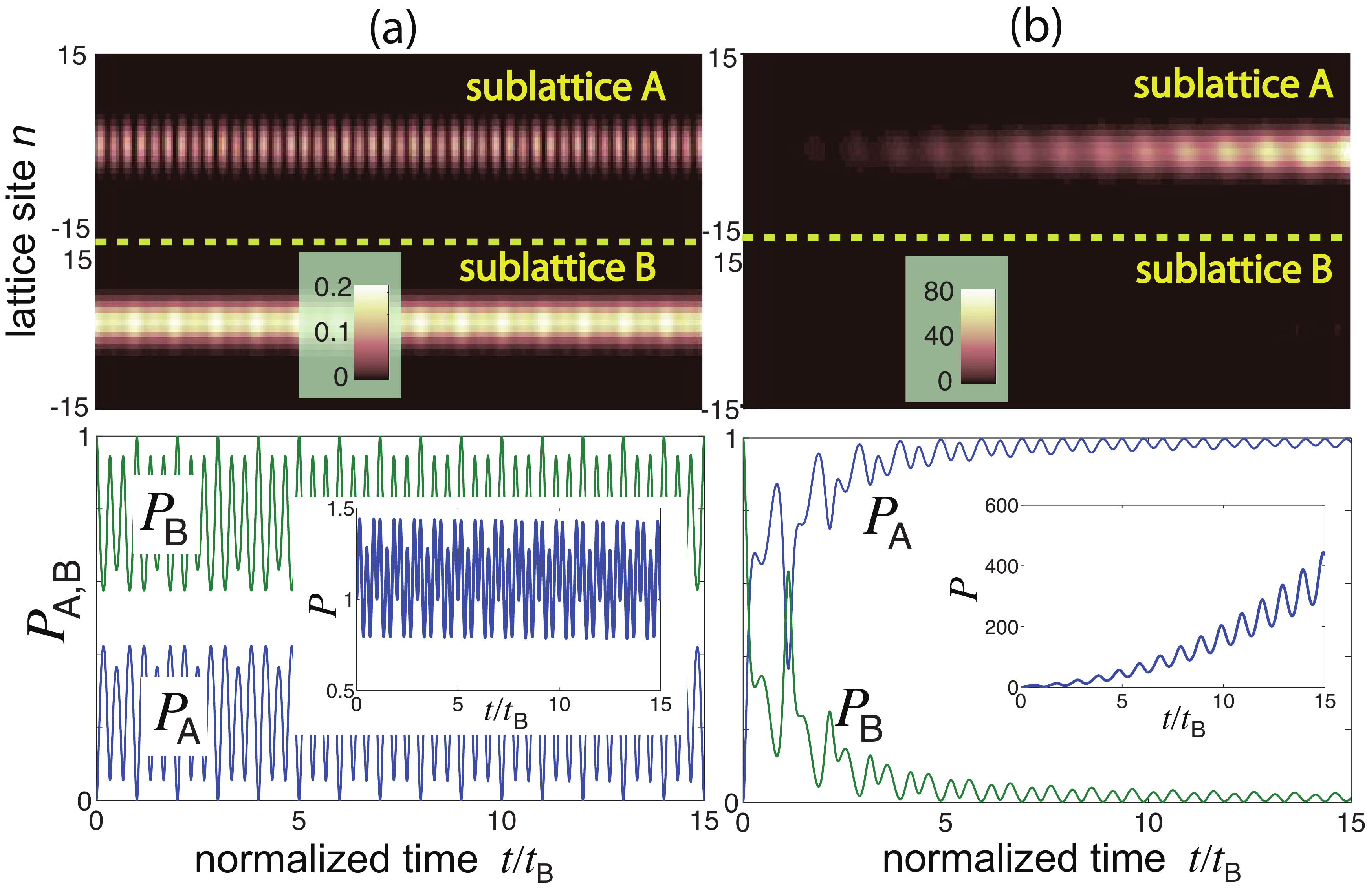}\\
   \caption{(color online) Chiral ZT in the lattice of Fig.1 under resonant driving (a) $F=-E_0$, and (b) $F=E_0$. Parameter values are as in Fig.2(c). At initial time only sublattice B is excited with a broad Gaussian wave packet of central momentum $k_0=0$, i.e. $b_n(0) \propto \exp[-(n/w)^2]$ with $w=4$ and $a_n(0)=0$. The upper panels show the evolution of $|a_n|^2$ and $|b_n|^2$ in a pseudo color map, whereas the lower panels depict the evolution of the normalized occupation probabilities $P_A$ and $P_B$ in the two sublattices A and B. The insets show the evolution of the norm $P(t)$. Note that in (a) tunneling into the sublattice A is weak and oscillatory, whereas in (b) irreversible ZT is observed with a secular growth of the norm $P(t)$. Time $t$ is normalized to the BO period $t_B= 2 \pi/F$.}
\end{figure}
 The usual picture of ZT in Hermitian crystals is that Eq.(14) describes a sequence of Landau-Zener transitions between Bloch bands as $k$ crosses the edges of the Brillouin zone, where the band gap is narrower \cite{r62b,r62c,rint}. The resulting quasi-periodic (oscillatory) dynamics is captured by the two Floquet exponents of Eq.(14), which are precisely the terms $ \pm F \theta /(2 \pi)$ that define the WS ladder energy spectrum [Eq.(10)]. Note that one can expand $H(k=Ft)$ in harmonics as 
$H(k=Ft)=H_0+\sum_{l=-q, l \neq 0}^{q} H_l \exp(-ilFt)$,
  where $H_l$ are constant $2 \times 2 $ matrices, and the eigenvalues of $H_0$ are precisely $\pm E_0$, given by Eq.(8). When $\rho_l=\theta_l=\varphi_l=0$ for $l<0$, i.e. when the non-Bloch bands collapse, $H(k=Ft)$ is composed solely by negative frequency components for $F>0$, or positive frequency components for $F<0$, so as, under near-resonant forcing, transitions preferentially occur from the $^{\prime}$lower$^{\prime}$ ($E_-$) to the $^{\prime}$higher$^{\prime}$ ($E_+$) energy bands for $F>0$, and viceversa for $F<0$ \cite{r67,r58}. According to the general properties of Floquet systems with one-sided harmonics \cite{r66}, provided that the force $F$ does not satisfy the resonance condition (13), the two Floquet exponents of Eq.(14) are given by $ \pm E_0$, the corresponding Floquet eigenstates are linearly independent and the Bloch dynamics is quasi periodic \cite{r58}. This result justifies the exactness of the WKB analysis [Eq.(13)] observed at the non-Bloch band collapse point [Fig.2(c)]. On the other hand, when $F$ satisfies the resonance condition (13), a Floquet EP arises \cite{r66}. In this case the dynamics ceases to be oscillatory and an irreversible tunneling from sublattice B (band $E_-$) to sublattice A (band $E_+$) is observed for $F>0$, but not for $F<0$, i.e. ZT becomes highly asymmetric under force reversal. In other words, chiral ZT is observed because of the appearance of a Floquet EP, which necessarily requires non-Bloch band collapse and resonant driving.  
  An example of chiral ZT in the lattice model of Fig.1(a) is shown in Fig.3. The figure depicts the dynamical evolution of the amplitudes $a_n(t)$ and $b_n(t)$ in physical space, norm $P(t)=\sum_n (|a_n|^2+|b_n|^2)$ and normalized occupation probabilities of the two sublattices A and B, $P_A(t)=(\sum_n|a_n|^2)/ P$ and $P_B(t)=(\sum_n|b_n|^2)/ P$, under resonance forcing. The initial condition corresponds to excitation of sublattice B. The figure clearly shows that irreversible ZT to sublattice A occurs for $F>0$, but not for $F<0$. In the former case irreversible tunneling is associated to a secular growth of the norm $P(t)$, which is a typical signature of a Floquet EP \cite{r66}, whereas in the latter case ZT probability is weak and oscillates with the characteristic BO period $t_B= 2 \pi /F$. For an imperfect non-Bloch band collapse, or for exact non-Bloch band collapse but non-resonant driving, the tunneling dynamics becomes oscillatory \cite{r58}. A simple physical explanation of the appearance of chirality in ZT for the lowest-order ($n=1$) resonance forcing at the Floquet EP is discussed in \cite{r58}.

{\it Conclusion.}  Non-Bloch band theory is a powerful tool to describe energy bands and to restore the bulk-boundary correspondence in non-Hermitian crystals showing the skin effect \cite{r22,r24,r32}. Here we have re-considered the old problem of Bloch oscillations and Zener tunneling in the framework of non-Bloch band theory, and unveiled a major physical effect unique to non-Hermitian systems, namely chiral Zener tunneling between dispersive Bloch bands. This phenomenon is observed under resonance forcing and is rooted in the collapse of non-Bloch bands and skin modes. Such results show major implications of non-Bloch band theory into coherent wave transport and are expected to be of broad relevance in different areas of physics, where wave transport is described by effective non-Hermitian models.

\newpage

\begin{widetext}
\begin{center}
\section*{ \bf Supplemental Material \\ for $^{\prime}$Non-Bloch band collapse and chiral Zener tunneling$^{\prime}$}
\end{center}
\end{widetext}
\renewcommand{\thesubsection}{S}
\renewcommand{\theequation}{S-\arabic{equation}}
\setcounter{equation}{0}
\begin{center}
{\it {\bf S.1. Non-Bloch band collapse}}\par
\end{center}
We consider a one-dimensional tight-binding lattice with two sites per unit cell, presented in the main text [Eq.(1)]. Here we show rather generally that, as the hopping amplitudes $\rho_l$, $\theta_l$ and $\varphi_l$ are one-sided, i.e. they vanish for either $l<0$ or $l>0$, the non-Bloch bands, defined by Eq.(6) given in the main manuscript [1], collapse to the two points $\pm E_0$ in complex energy plane, where
\begin{equation}
E_0= \sqrt{\rho_0^2+\varphi_0 \theta_0}.
\end{equation}
Likewise, the generalized Brillouin zone $\tilde{C}_{\beta}$ shrinks toward the point $\beta=0$ (for positive one-sided hoppings) or  $1/ \beta=0$ (for negative one-sided hoppings), indicating that simultaneous collapse of skin modes occurs.\\ 
For the sake of definiteness, we will consider the case of positive one-sided hopping amplitudes, i.e. $\rho_l=\varphi_l=\theta_l \sim 0$ for $l<0$, however a similar analysis holds {\em mutatis mutandis} for the negative one-sided hopping amplitudes.\\
Let us indicate by $\beta_1(E)$, $\beta_2(E)$,..., $\beta_{4q}(E)$ the $4q$ roots of the algebraic equation $Q(\beta)=E^2$, i.e 
\begin{equation}
\beta^{2q} \left\{ P_{2q}(\beta)-E^2 \right\}+R_{2q-1}(\beta)=0
\end{equation}
ordered such that $|\beta_1(E)| \leq |\beta_2(E)| \leq ... \leq |\beta_{4q}(E)|$. Then the energies $E$ forming the non-Bloch bands, i.e. the energy spectrum of the system with OBC, are obtained from the implicit relation 
\begin{equation}
|\beta_{2q}(E)|=|\beta_{2q+1}(E)|
\end{equation}
which also defines the generalized Brillouin zone $\tilde{C}_{\beta}$ [1].  In Eq.(S-2) we have set
\begin{eqnarray}
P_{2q}(\beta) & = & \sum_{n=0}^{2q} \left( \sum_{l} (\rho_l \rho_{n-l} + \theta_l \varphi_{n-l}) \right) \beta^n \\
R_{2q-1}(\beta) & = & \sum_{n=-2q}^{-1} \left(  \sum_{l} (\rho_l \rho_{n-l} +\theta_l \varphi_{n-l})  \right) \beta^{n+2q} \nonumber
\end{eqnarray}
and $\beta \equiv \exp(-ik)$. Let us assume now that the hopping $\rho_l$, $\theta_l$ and $\varphi_l$ for $l<0$ are vanishing, say small parameters of order $\sim \epsilon$. Then from an inspection of Eq.(S-4) it readily follows that the coefficients of the polynomial $R_{2q-1}(\beta)$ are small and of order $\sim \epsilon$. On the other hand, the coefficients of polynomial $P_{2q}(\beta)$ are of order $\sim \epsilon^0$. To provide an asymptotic form of the roots to Eq.(S-2), we have to distinguish two cases.\\
(i) The energy $E$ is far from $ \pm E_0$. Let us assume that $|E^2-E_0^2|$ is much larger than $\sim \epsilon$. Taking into account that $P_{2q}(\beta=0)=E_0^2+O(\epsilon)$, it is clear that, at leading order, Eq.(S-2) is satisfied for either 
\begin{equation}
P_{2q} (\beta) \simeq E^2
\end{equation}
or for large $\beta$, $\beta \sim O(\epsilon^{1/2q})$, namely
\begin{equation}
\beta^{2q} \simeq \frac{R_{2q-1}(0)}{E^2-E_0^2} \simeq \frac{\rho_0 \rho_{-2q} + \theta_0 \varphi_{-2q}}{E^2-E_0^2}.
\end{equation}
Hence, the $2q$ roots of Eq.(S-6), of order $\sim \epsilon^{1/(2q)}$, are the roots $\beta_1(E)$, $\beta_2(E)$, ..., $\beta_{2q}(E)$, whereas the other $2q$ roots of Eq.(S-5), of order $\sim \epsilon^0$, are the other roots $\beta_{2q+1}(E)$, $\beta_{2q+2}(E)$, ..., $\beta_{4q}(E)$. As a consequence, the condition (S-3) can never be satisfied, since $\beta_{2q}(E)$ is small (of order $\sim \epsilon^{1/2q}$) whereas $\beta_{2q+1}(E)$ is of order $\sim \epsilon^0$. This means that, as $\epsilon$ goes to zero, the allowed energies $E^2$, defining the non-Bloch bands, should collapse to the value $E_0^2$.\\
(ii) For $|E^2-E_0^2| \sim \epsilon$, let us write Eq.(S-2) in the equivalent form
\begin{equation}
\beta^{2q+1} \tilde{P}_{2q-1}(\beta)+\tilde{R}_{2q}(\beta)=0
\end{equation}   
where the polynomials $\tilde{P}_{2q-1}(\beta)$ and $\tilde{R}_{2q-1}(\beta)$, respectively of order $(2q-1)$ and $2q$, are given by
\begin{eqnarray}
\tilde{P}_{2q-1}(\beta) & = &  \frac{1}{\beta} \left\{ P_{2q}(\beta)-P_{2q}(0) \right\} \\
\tilde{R}_{2q}(\beta) & = & R_{2q-1}(\beta) + [P_{2q}(0)-E^2] \beta^{2q}.
\end{eqnarray}
Note that, since $P_{2q}(0)-E^2 \sim E_0^2-E^2 \sim \epsilon$, the coefficients of the polynomial $\tilde{R}_{2q}(\beta)$ are small of order $\sim \epsilon$, whereas the coefficients of the polynomial $\tilde{P}_{2q-1}(\beta)$ are of order $\sim \epsilon^0$. Therefore, Eq.(S-7) is satisfied for either 
\begin{equation}
\tilde{P}_{2q-1}(\beta) \simeq 0
\end{equation}
 or for
\begin{equation}
\beta^{2q+1}{P}_{2q-1}(0)+ \tilde{R}_{2q}(0) \simeq 0.
\end{equation}
 The $(2q-1)$ roots to Eq.(S-10) define the branches $\beta_{2q+2}(E)$, ...$\beta_{4q}(E)$, which are of order $\sim \epsilon^0$, whereas the $(2q+1)$ roots to Eq.(S-11), which are small and of order $\sim \epsilon^{1/(2q+1)}$, define the branches $\beta_{1}(E)$, $\beta_{2}(E)$,..., $\beta_{2q+1}(E)$ (note that at leading order such roots are almost independent of energy $E^2 \simeq E_0^2$). In this case the implicit equation (S-3), which determines the OBC energy spectrum, can be satisfied being $|\beta_{2q}|$ and $|\beta_{2q+1}|$ small and of the same order $\sim \epsilon^{1/(2q+1)}$. To conclude, as $\epsilon \rightarrow 0$, the OBC energy spectrum $E^2$ shrinks to the value $E_0^2$, defined by Eq.(S-1), and the generalized Brillouin zone $\tilde{C}_{\beta}$, defined by the loci  $\beta$ in complex plane satisfying the condition (S-3), shrinks toward the origin $\beta=0$ with a convergence of order $\sim \epsilon^{1/(2q+1)}$, where $q$ is the most distant site where hopping in the chain can occur. This means that, at the non-Bloch band collapse, the skin modes also shrink toward the right edge sites $a_N$ and $b_N$ of the lattice, with vanishing occupation of all other sites of the chain.\\
 It is worth providing a different and more direct proof of non-Bloch  band collapse and simultaneous skin mode coalescence in lattices with one-sided hopping. To this aim, let us notice that, for systems with OBC, the allowed energies $E$ are the eigenvalues of the linear system of equations
 \begin{eqnarray}
 (E-\rho_0)a_n- \theta_0 b_n & = & \sum_{l \neq n} \left( \rho_{n-l} a_l + \theta_{n-l} b_l \right) \\
 (E+\rho_0)b_n- \varphi_0 a_n & = & \sum_{l \neq n} \left(- \rho_{n-l} b_l + \varphi_{n-l} a_l \right) \;\;\;\;
 \end{eqnarray}
 ($n=1,2,3,...,N$) with the following boundary conditions
\begin{equation}
a_n=b_n=0
\end{equation}
for any $n \leq 0$ and $n \geq N+1$, where $N$ is the number of unit cells of the crystal. Rather generally, such a boundary-value problem yields $2N$ distinct eigenvalues for the energy $E$ and corresponding linearly-independent eigenvectors, that define the non-Bloch energy spectrum and skin modes of the lattice with OBC. Let us now assume, like in the previous analysis, a lattice with one-sided positive hopping, i.e. $\rho_l=\theta_l=\varphi_l=0$ for $l<0$. This means that the sums on the right hand side of Eqs.(S-12) and (S-13) are restricted to $0<l<n$.
If we consider Eq.(S-12) and (S-13) for $n=1$, taking into account that $a_l=b_l=0$ for $l \leq 0$ it readily follows that 
\begin{eqnarray}
 (E-\rho_0)a_1- \theta_0 b_1 & = & 0 \\
 (E+\rho_0)b_1- \varphi_0 a_1 & = & 0.
\end{eqnarray}
The above equations can be satisfied by taking  either $a_{1}=b_{1}=0$, or $a_{1}, b_{1} \neq 0$ with the energy $E$ satisfying the determinantal equation
\begin{equation}
\left|
\begin{array}{cc}
E-\rho_0 & - \theta_0 \\
-\varphi_0 & E+ \rho_0
\end{array}
\right|=0,
\end{equation}
i.e. $E= \pm E_0$. The latter choice, however, is forbidden. In fact, writing Eqs.(S-12) and (S-13)  for $n=2$ yields
\begin{eqnarray}
 (E-\rho_0)a_{2}- \theta_0 b_{2} & = &  \rho_1 a_1+\theta_1 b_1 \\
 (E+\rho_0)b_{2}- \varphi_0 a_{2} & = & -\rho_1 b_1+\varphi_1 a_1  
\end{eqnarray}
which can not be rather generally solved because of the determinantal condition (S-17). Therefore we should assume $a_1=b_1=0$. The above reasoning can be iterated to show that $a_n=b_n=0$ for any $n=1,2,...,N-1$. Clearly, for $n=N$ one obtains
\begin{eqnarray}
 (E-\rho_0)a_N- \theta_0 b_N & = & 0 \\
 (E+\rho_0)b_N- \varphi_0 a_N & = & 0.
\end{eqnarray}
Since this is the last unit cell of the lattice, we can now assume $a_N,b_N \neq 0$ provided that the energy $E$ satisfies the determinantal equation (S-17), i.e. $E= \pm E_0$. In such a case the amplitudes $a_N$ and $b_N$ are related one another by the condition $( \pm E_0-\rho_0)a_N= \theta_0 b_N$. To conclude, for one-sided hopping, the $2N$ eigenenergies of the boundary-value linear problem, defined by Eqs.(S-12, S-13,S-14), collapse to the two allowed values $\pm E_0$, and there are only two linearly-independent eigenvectors, corresponding to non-vanishing occupation of the right-edge unit cell $n=N$ of the lattice. This means that non-Bloch band collapse is also associated to the coalescence of corresponding eigenstates, i.e. the two energies $E= \pm E_0$ are two EPs (each of order $N$) of the OBC spectral problem. 
 \begin{center}
{\it {\bf S.2. Bloch dynamics under a dc force: Wannier-Stark ladders}}\par
\end{center}
Bloch oscillations and Zener tunneling between the two dispersive Bloch bands, induced by the external force $F$, are at best captured by considering wave dynamics in the Bloch basis representation. Here we consider an infinitely-extended lattice. After setting
\begin{eqnarray}
A(k,t) & = &  \frac{1}{2 \pi} \sum_{n} a_n(t) \exp (-ikn)  \\ 
 B(k,t ) & = & \frac{1}{2 \pi} \sum_{n} b_n(t) \exp (-ikn),
\end{eqnarray}
 the evolution equation for the spectral amplitudes  $A$ and $B$ is governed by the coupled equations
\begin{equation}
i \frac{\partial}{\partial t}
\left(
\begin{array}{c}
A \\
B
\end{array}
\right)= \left( H(k)-iF \frac{\partial}{\partial k} \right)
\left(
\begin{array}{c}
A \\
B
\end{array}
\right)
\end{equation}
where $H(k)$ is the Hamiltonian in the Bloch basis representation [Eq.(3) of the main manuscript].
Once the spectral equation (S-24) is solved, the occupation amplitudes of the various lattice sites in the Wannier basis representation are then obtained from the inverse relations
\begin{eqnarray}
a_n(t) & = & \int_{-\pi}^{\pi} dk A(k,t) \exp(ikn) \\
 b_n(t)& = & \int_{-\pi}^{\pi} dk B(k,t) \exp(ikn).
\end{eqnarray}
To calculate the energy spectrum and the formation of two WS ladders, we look for a solution to Eq.(S-24) of the form 
\begin{equation}
\left(
\begin{array}{c}
A(k,t) \\
B(k,t)
\end{array}
\right)= \left(
\begin{array}{c}
\mathcal{A}(k) \\
\mathcal{B}(k)
\end{array}
\right) \exp(-i Et+ikE/F)
\end{equation}
where $E$ is the eigenenergy. The spectral amplitudes $\mathcal{A}(k)$, $\mathcal{B}(k)$ satisfy the coupled equations
\begin{equation}
i F \frac{d}{dk}
\left(
\begin{array}{c}
\mathcal{A} \\
\mathcal{B}
\end{array}
\right)= H(k) \left(
\begin{array}{c}
\mathcal{A} \\
\mathcal{B}
\end{array}
\right)
\end{equation}
which can be formally solved, from $k=-\pi$ to $k= \pi$, yielding  
\begin{equation}
\left(
\begin{array}{c}
\mathcal{A} ( \pi) \\
\mathcal{B} ( \pi)
\end{array}
\right)=
U
\left(
\begin{array}{c}
\mathcal{A} (-\pi) \\
\mathcal{B} (-\pi)
\end{array}
\right)
\end{equation}
where $U$ is the ordered exponential matrix given by Eq.(11) in the main manuscript, i.e.
\begin{equation}
U= \int_{-\pi}^{ \pi} dk \exp \left\{ -i H(k) /F \right\}.
\end{equation}
Note that, for construction [Eqs.(S-22) and (S-23)], the amplitudes $A(k,t)$ and $B(k,t)$ are periodic in $k$ with $ 2 \pi$ period. This means that the spectral amplitudes $\mathcal{A}(k)$ and $\mathcal{B}(k)$ should satisfy the following boundary conditions
\begin{equation}
\left(
\begin{array}{c}
\mathcal{A} (\pi) \\
\mathcal{B}(\pi)
\end{array}
\right) =\exp(-2  \pi i E/F) \left(
\begin{array}{c}
\mathcal{A} (- \pi) \\
\mathcal{B}(- \pi)
\end{array}
\right) 
\end{equation}
A comparison of Eqs.(S-29) and (S-31) shows that $(\mathcal{A}(-\pi), \mathcal{B}(-\pi))^T$ should be an eigenvector of the matrix $U$ and $\exp(-2 \pi i E/F)$ the corresponding eigenvalue. Taking into account that $\det U=U_{11}U_{22}-U_{12}U_{21}=1$ [this is because the trace of $H(k)$ vanishes], the eigenvalues $\lambda_{1,2}$ of $U$ are given by $\lambda_{1,2}= \exp( \pm i \theta)$, where the complex angle $\theta$ is defined by the relation
\begin{equation}
\cos \theta = \frac{1}{2} {\rm Tr} (U)=\frac{U_{11}+U_{22}}{2}.
\end{equation}
The boundary condition (S-31) is thus satisfied for energies $E=E_l$ satisfying the condition
\begin{equation}
2 \pi E_l/F= \pm \theta + 2 l \pi
\end{equation}
where $l$ is an arbitrary integer number. From Eq.(S-33) one then follows Eq.(10) given in the main text for the energies $E^{(WS)}_{\pm}$ of the two WS ladders. The corresponding $l$-th eigenstate in the Wannier basis is obtained from the Fourier integrals Eqs.(S-25) and (S-26), i.e.
\begin{eqnarray}
a_n^{(l)}(t) & = &  \exp(-iE_l t)  \times \\
&  \times & \int_{-\pi}^{\pi} dk \mathcal{A}(k) \exp \left\{ \pm i \frac{k \theta}{2 \pi} +ik(l+n) \right\} \nonumber
\end{eqnarray}
and a similar expression for $b_n^{(l)}(t)$. Note that, as the WS index $l$ is varied by $\pm 1$, the WS eigenstate in the Wannier basis representation just shifts along the lattice by one unit cell, either forward or backward. Moreover, 
 since the spectral amplitude $\mathcal{A}(k)$ is continuous with all its $k$-derivatives in the interval $(-\pi,\pi)$, the amplitude  $a_n^{(l)}$, for a fixed mode index $l$, decays as $n \rightarrow \pm \infty$ faster than any power-law $ \sim 1/|n|^p$ ($p$ arbitrarily large). This means that, even in a system with OBC,  the localization induced by the external force, i.e. the formation of WS ladders, overcomes the skin effect to squeeze the modes toward one edge.\\
\\  
\begin{figure}
\includegraphics[width=84mm]{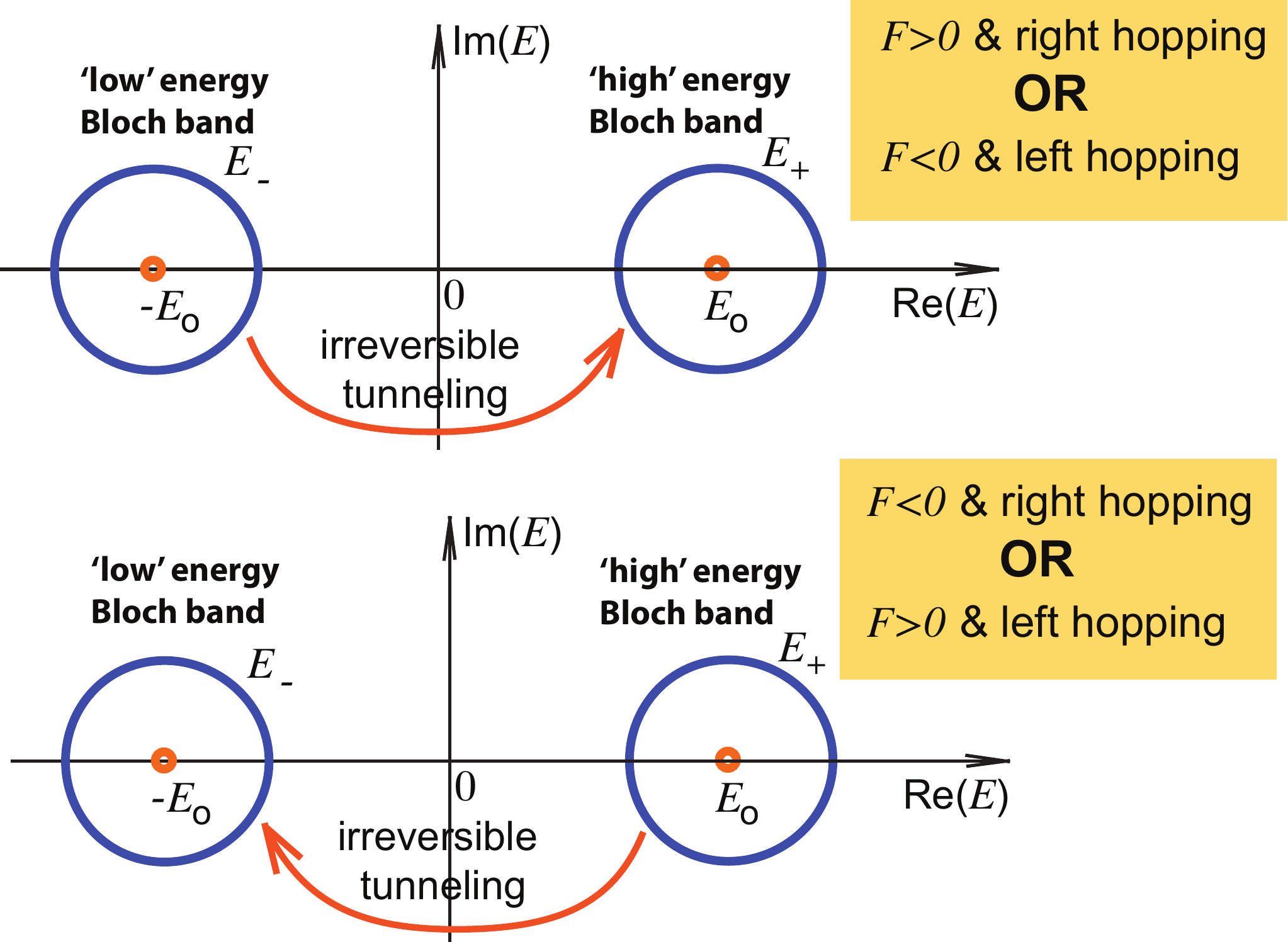}
\caption{\small (Color online) Chiral ZT between the two dispersive Bloch bands, depending on the direction of the force $F$ and of hopping in the lattice. Right hopping corresponds to $\rho_l=\theta_l=\varphi_l=0$ for $l<0$, whereas left hopping indicates the case $\rho_l=\theta_l=\varphi_l=0$ for $l>0$.}
\end{figure}

 \begin{center}
{\it {\bf S.3. Zener tunneling under off-resonance forcing and near the non-Bloch band collapse point}}\par
\end{center}
The chiral nature of ZT between the two dispersive Bloch bands, induced by the external force and discussed in the main text, is observed when the force satisfies the resonance condition $F= \pm 2 E_0/n$, with $n$ an integer number, {\em and} the two non-Bloch bands collapse. In this case a Floquet EP is found and irreversible (i.e. non-oscillatory) tunneling from the $^{\prime}high^{\prime}$ energy dispersive Bloch band $E_+$ to the  $^{\prime}low^{\prime}$ energy dispersive Bloch band $E_-$, or viceversa, is observed, depending on the direction of the force $F$ and of hopping in the lattice, as schematically shown in Fig.4.\\
 When either one of the two above conditions is not met, the ZT dynamics becomes oscillatory (similarly to an Hermitian lattice), i.e. the irreversible (chiral) tunneling from one band to the other one disappears. Here we provide some examples of oscillatory dynamics under off-resonance forcing or near the non-Bloch band collapse point.\\  
\\
{\it Off-resonance ZT dynamics.} For off-resonance driving, ZT between the two bands becomes oscillatory, even though the two non-Bloch bands collapse.  In fact, for off-resonance driving the two interleaved  WS energy ladders do not coalesce, and the dynamical evolution of occupation amplitudes $a_n(t)$ and $b_n(t)$ in the lattice, for a given initial condition, is quasi-periodic with two characteristic time periods determined by the spacing between the spectral modes within the same WS ladder and the spectral distance between the two WS ladders. Such time periods are the BO period $t_B= 2 \pi /F$ and the inter-WS period $t_{WS}=  (\pi / \theta) t_B$, respectively. An example of oscillatory ZT for off-resonance forcing is shown in Fig.5. Parameter values and initial excitation of the lattice are the same as in Fig.3 of the main text, except that the force $F$ is set at $F= \pm 0.7 E_0$. Clearly, for both $F>0$ and $F<0$ ZT is oscillatory and successive revivals of the initial excitation condition (due to quasi-periodicity) is clearly observed, regardless of the direction of the force.\\
\\ 
{\it ZT near the non-Bloch band collapse.} If the two non-Bloch bands do not exactly collapse, i.e. $\delta \neq \pm t$, the WS ladders do not coalesce (Fig.2 in the main text) and, like for the off-resonance driving case discussed above, the ZT dynamics becomes oscillatory with two periodicities, the long-period  of the dynamics, $t_{WS}$, diverging as the non-Bloch band collapse point is attained. A typical example of ZT dynamics for resonance forcing $F= \pm E_0$ and close to the band collapse point is shown in Fig.6.\\
\\
\\  
\begin{figure}
\includegraphics[width=84mm]{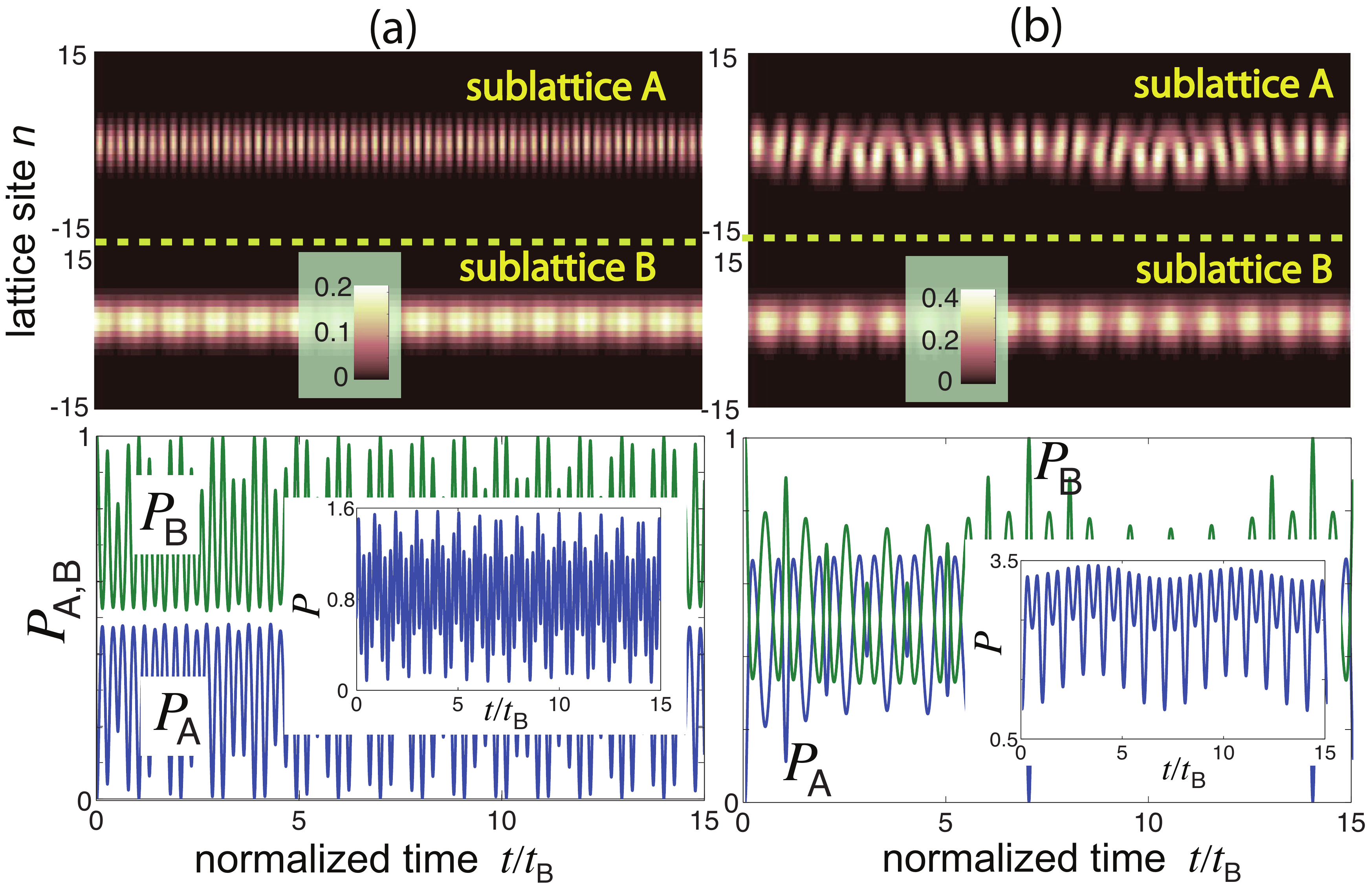}
\caption{\small (Color online) Bloch dynamics and ZT under off-resonance forcing. Parameter values are as in Fig.3 of the main text, except that $F=-0.7 E_0$ in (a), and $F=0.7 E_0$ in (b).}
\end{figure}
 \begin{figure}
\includegraphics[width=84mm]{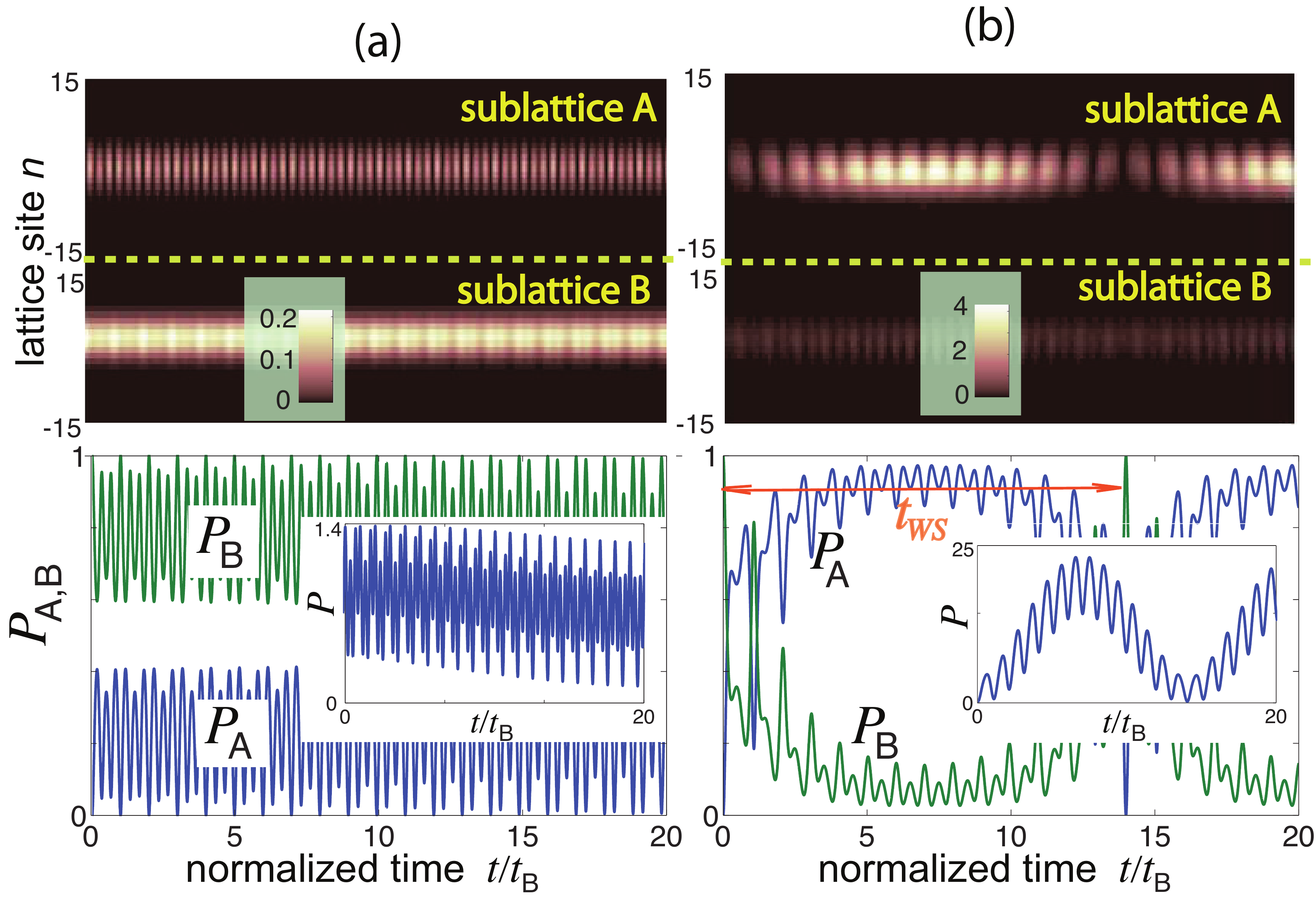}
\caption{\small (Color online) Bloch dynamics and ZT near the non-Bloch band collapse point. Parameter values are as in Fig.3 of the main text, except that $\delta=0.92 t$. (a) $F=-E_0$, and (b) $F=E_0$.}
\end{figure}
{\it {\bf S.4. A simple physical picture of chiral Zener tunneling}}\par
To provide simple physical insights into the occurrence of chiral ZT, let us consider the model shown in Fig.1(a) of the main manuscript. 
For a force $F$ close to the first resonance tongue, i.e. for $F =2 E_0 \simeq 2 \Delta$, irreversibility of ZT between the weakly-dispersive Bloch bands can be readily explained either in the Bloch band domain  or 
in the real-space domain.\\
\\
{\it Bloch-band domain}. Let us specialize Eq.(14) given in the main text, describing BOs and ZT between the two dispersive Bloch bands, for the model shown in Fig.1(a).
The Hamiltonian for this model reads
\[
H(k)= \left(
\begin{array}{cc}
\Delta & t_0+(t+\delta) \exp(-ik) \\
 t_0+(t-\delta) \exp(ik) & - \Delta
\end{array}
\right).
\]
\begin{figure}
\caption{\small (Color online) Schematic of the binary lattice of Fig.1(a) in the undriven ($F=0$) and resonantly-driven (first resonance tongue $F= \pm 2 \Delta$) cases. The dashed curves show the tunneling paths, described by the hopping rates $\theta_0=\varphi_0=t_0$ and $\theta_1=t+ \delta$, $\varphi_{-1}=t-\delta$. The thin solid curves show possible non-nearest neighbor interaction $\theta_{-1}$, $\varphi_{1}$ between the two sublattices, which is not considered in the model of Fig.1(a). In the driven case, sites in adjacent unit cells acquire a potential energy shift $F$ due to the external force. Note that for $F=2 \Delta$ (lower panel) the sites in sublattices A and B, belonging to adjacent unit cells, are in resonance and thus tunneling between dimers is allowed. The tunneling is unidirectional for $\delta= \pm t$: for $\delta=t$, tunneling is allowed only from B to A sublattices, wheres for $\delta=-t$ only from A to B sublattices. When non-nearest neighbor hopping are considered, chiral tunneling is also possible for $F=-2\Delta$.}
\end{figure}
 After setting  $f_{A}(t)=g_A(t) \exp(-i \Delta t)$,  $f_{B}(t)=g_B(t) \exp(i \Delta t)$, the evolution equations for the amplitude probabilities to find the particle in either sublattices A and B read
\begin{eqnarray}
i \frac{dg_A}{dt} & = & \left\{ t_0 \exp(2i \Delta t)  +(t+\delta) \exp(2i \Delta t-iFt) \right\} g_B  \;\;\;\;\;\;\;\;  \\
i \frac{dg_B}{dt} & = & \left\{ t_0 \exp(-2i \Delta t) +(t-\delta) \exp(-2i \Delta t+iFt) \right\} g_A \;\;\;\; \;\;\;\;\;\;\;\;  
\end{eqnarray}
We consider the weak-dispersive band limit $\Delta \gg t_0, t, |\delta |$ and focus on the first resonance tongue [$n=1$ in Eq.(13)], i.e. we assume $F$ close to $ \pm 2E_0 \simeq \pm 2\Delta$. For higher-order resonances, multiple time scale analysis would be in order to explain resonant tunneling, however the main physics underlying chiral ZT is fully captured by considering the first resonance tongue. In such a limit, in the rotating-wave approximation we can disregard rapidly-oscillating terms on the right hand side of Eqs.(S-35) and (S-36). For $F<0$, all terms are rapidly oscillating and thus, in the rotating-wave approximation, the external force does not induce transitions between the two dispersive Bloch bands, regardless of the value of the non-Hermitian parameter $\delta$. In a higher-order approximation, one observes small and oscillatory ZT transitions (see e.g. Fig.3(a) in the main manuscript and Figs. S1 and S2). On the other hand, for $F>0$ one obtains
\begin{eqnarray}
i \frac{dg_A}{dt} & = &(t+\delta) \exp(-i \Omega t)  g_B   \\
i \frac{dg_B}{dt} & = & (t-\delta) \exp(i \Omega t)  g_A 
\end{eqnarray} 
where we have set $\Omega \equiv F-2 \Delta$. For $\Omega \neq 0$ (off-resonance ZT), regardless o the value of $\delta$ the dynamics described by Eqs.(S-37) and (S-38) is oscillatory. The same scenario occurs for resonance forcing $\Omega=0$ and $\delta \neq \pm t$, i.e. far from the non-Bloch band collapse point. On the other hand, for resonance forcing $\Omega=0$ and $\delta=t$, Eqs.(S-37) and (S-38) show that, if the system is initially prepared with excitation in sites B, one has $g_B(t)=g_B(0)$ and $g_A(t)=2 t g_B(0) t$, i.e. a secular growth of excitation in A is observed [Fig.3(b) in the main manuscript]. This regime corresponds to chiral ZT and to a Floquet EP of the time-periodic system (Eq.(14) in the main text)  [2].\\
Finally, it should be noted that if the model of Fig.1(a) is extended to include non-nearest neighbor hopping amplitudes $\theta_1$ and $\varphi_{-1}$, chiral ZT can be observed also for $F=-2 \Delta$ resonance driving. In fact, for $F \simeq -2 \Delta$, in the rotating-wave approximation the coupled equations for the amplitudes $g_{A}$ and $g_B$ read
\begin{eqnarray}
i \frac{dg_A}{dt} & = &\theta_{-1} \exp(i \Omega t)  g_B   \\
i \frac{dg_B}{dt} & = & \varphi_{1} \exp(-i \Omega t)  g_A 
\end{eqnarray}
where we have set $\Omega \equiv F+2 \Delta$. For resonant driving $\Omega=0$, irreversible tunneling, from sublattice B to sublattice A, is thus observed for $\varphi_1=0$, $\theta_{-1} \neq 0$, whereas irreversible tunneling from sublattice A to sublattice B is observed for $\varphi_1 \neq 0$, $\theta_{-1}= 0$, according to the general scenario shown in Fig.4.\\
\\
{\it Real-space domain}. Figure 7 schematically shows the binary lattice for model of Fig.1(a) in the undriven ($F=0$) and resonantly-driven ($F= \pm 2 \Delta$) cases. The vertical displacements of the sites in the lattice depict the site-energy potential, which is affected by the external force $F$. In the weakly dispersive band limit $\Delta \gg t, t_0 |\delta|$, tunneling between adjacent sites is weak in the undriven $F=0$ and resonant driving $F=-2 \Delta$ cases, because adjacent sites in the lattice are out of resonance. On the other hand, for the resonant forcing $F= 2 \Delta$ isolated dimers in the two sublattices A and B are set in resonance by the external force, as shown in the bottom panel of Fig.7. Thus ZT between the two sublattices is allowed in this regime, and rather generally it is oscillatory, i.e. excitation is periodically transferred from sublattices A and B. However, for $\delta=\pm t$ tunneling becomes unidirectional, namely for $\delta=t$ tunneling can arise from sublattice B to sublattice A, whereas for $\delta=-t$ tunneling can arise from sublattice A to sublattice B, accordion go the general scenario shown in Fig.4. Therefore, irreversible excitation of sublattices A (or B) is observed at $\delta=t$ (or $\delta=-t$), corresponding to chiral ZT.\\ 
Finally, it should be noted that, if the model of Fig.1(a) is extended to include non-nearest neighbor hopping $\varphi_{1}$, $\theta_{-1}$ (indicated by the thin solid lines in Fig.7), chiral ZT could be observed also for $F=-2\Delta$, with the direction of irreversible tunneling ruled as in Fig.4.\\
\small
\\
\noindent
{\bf References}\\
{[1]} K. Yokomizo and S. Murakami, Non-Bloch Band Theory for Non-Hermitian Systems, Phys. Rev. Lett. {\bf 123}, 066404 (2019).\\
{[2]} S. Longhi, Floquet exceptional points and chirality in non-Hermitian Hamiltonians, J. Phys. A {\bf 50}, 505201 (2017).\\

\end{document}